\newcommand{\nc}{\newcommand}
\nc{\kms}{\,{km\,s$^{-1}$}}
\nc{\sgra}{Sgr\ts A}
\nc{\sgrastar}{Sgr\ts$\rm {A}^{*}$}
\nc{\sgraeast}{Sgr\ts A\ts East}
\nc{\sgrawest}{Sgr\ts A\ts West}
\nc{\sgracomp}{Sgr\ts A\ts Complex}
\nc{\as}{\arcsec}
\nc{\HI}{H\,{\sc i}}
\nc{\HII}{H\,{\sc ii}}
\nc{\CI}{C\,{\sc i}}
\nc{\hto}{H$_{2}$O}
\nc{\htmo}{H$_{2}^{16}$O}
\nc{\htio}{H$_{2}^{18}$O}
\nc{\ciso}{C$^{18}$O}

\documentclass{aa}

\usepackage{graphicx}
\usepackage{txfonts}
\input{psfig}

\begin{document}

\title{Odin observations of \hto\  in the Galactic Centre\thanks{Based
on observations with Odin, a Swedish-led satellite project funded
  jointly by the Swedish National Space Board (SNSB), the Canadian
  Space Agency (CSA), the National Technology Agency of
  Finland (Tekes) and Centre National d'Etude Spatiale (CNES). The
  Swedish Space Corporation was the industrial prime contractor
  and is also responsible for the satellite operation.}}

\author{Aa.\,Sandqvist\inst{1}
  \and P.\,Bergman\inst{2}
  \and J.H.\,Black\inst{2}
  \and R.\,Booth\inst{2}
  \and V.\,Buat\inst{3}
  \and C.L.\,Curry\inst{4}
  \and P.\,Encrenaz\inst{5}
  \and E.\,Falgarone\inst{6}
  \and P.\,Feldman\inst{7}
  \and M.\,Fich\inst{4}
  \and H.G.\,Floren\inst{1}
  \and U.\,Frisk\inst{8}
  \and M.\,Gerin\inst{6}
  \and E.M.\,Gregersen\inst{9}
  \and J.\,Harju\inst{10}
  \and T.\,Hasegawa\inst{11}
  \and \AA .\,Hjalmarson\inst{2}
  \and L.E.B.\,Johansson\inst{2}
  \and S.\,Kwok\inst{11}
  \and B.\,Larsson\inst{1}
  \and A.\,Lecacheux\inst{12}
  \and T.\,Liljestr\"om\inst{13}
  \and M.\,Lindqvist\inst{2}
  \and R.\,Liseau\inst{1}
  \and K.\,Mattila\inst{10}
  \and G.F.\,Mitchell\inst{14}
  \and L.\,Nordh\inst{15}
  \and M.\,Olberg\inst{2}
  \and A.O.H.\,Olofsson\inst{2}
  \and G.\,Olofsson\inst{1}
  \and L.\,Pagani\inst{5}
  \and R.\,Plume\inst{11}
  \and I.\,Ristorcelli\inst{16}
  \and F.v.\,Sch\'eele\inst{8}
  \and G.\,Serra\inst{16}
  \and N.F.H.\,Tothill\inst{14}
  \and K.\,Volk\inst{11}
  \and C.D.\,Wilson\inst{9}
  \and A.\,Winnberg\inst{2} 
}

\institute{Stockholm Observatory, SCFAB-AlbaNova, SE-106 91 Stockholm,
  Sweden\\ \email{aage@astro.su.se}
  \and Onsala Space Observatory, SE-439 92 Onsala, Sweden
  \and Laboratoire d'Astronomie Spatiale, BP 8, FR-13376 Marseille
  CEDEX 12, France
  \and Department of Physics, University of Waterloo, Waterloo, ON N2L
  3G1, Canada
  \and LERMA \&\  FRE 2460 du CNRS, Observatoire de Paris, 61, ave de
  l'Observatoire,, FR-75014 Paris, France 
  \and LERMA \&\  FRE 2460 du CNRS, Ecole Normale Sup\'erieure, 24 rue
  Lhomond, FR-75005 Paris, France
  \and Herzberg Institute of Astrophysics, 5071 West Saanich Road,
  Victoria, BC V9E 2E7, Canada
  \and Swedish Space Corporation, P. O. Box 4207, SE-171 04 Solna,
  Sweden
  \and Department of Physics and Astronomy, McMaster University,
  Hamilton, ON L8S 4M1, Canada
  \and Observatory, P. O. Box 14, University of Helsinki, FI-00014
  Helsinki, Finland
  \and Department of Physics and Astronomy, University of Calgary,
  Calgary, AB T2N 1N4, Canada
  \and LESIA, Observatoire de Paris, Section de Meudon, 5, Place Jules
  Janssen, FR-92195 Meudon CEDEX, France
  \and Mets\"ahovi Radio Observatory, Helsinki University of
  Technology, Otakaari 5A, FI-02150 Espoo, Finland 
  \and Department of Astronomy and Physics, Saint Mary's University,
  Halifax, NS B3H 3C3, Canada
  \and Swedish National Space Board, Box 4006, SE-171 04 Solna,
  Sweden
  \and CESR, 9 Avenue du Colonel Roche, B.P. 4346, FR-31029 Toulouse, France}

\offprints{\\ Aage Sandqvist, \email{aage@astro.su.se}}

\date{Received $<$date$>$; accepted $<$date$>$}

\abstract{The Odin satellite has been used to detect emission and
  absorption in the 557-GHz \htmo\  line in the Galactic Centre
  towards the \sgrastar\ Circumnuclear Disk (CND), and the \sgra\ +20
  \kms\ and +50 \kms\ molecular clouds. Strong broad \hto\  emission
  lines have been detected in all three objects. Narrow
  H$_{2}$O absorption lines are present at all three positions and
  originate along the lines of sight in the 3-kpc Spiral
  Arm, the $-$30 \kms\ Spiral Arm and the Local Sgr Spiral Arm. Broad
  \hto\  absorption lines near $-130$ \kms\ are also observed,
  originating in the Expanding Molecular Ring. A new molecular feature
  (the ``High Positive Velocity Gas'' - HPVG) has been identified in the
  positive velocity range of $\approx$\  +120 to +220 \kms, seen
  definitely in absorption against the stronger dust continuum emission from
  the +20 \kms\ and +50 \kms\ clouds and possibly in emission towards
  the position of \sgrastar\ CND. The 548-GHz \htio\  isotope line
  towards the CND is not detected at the 0.02 K (rms) level.  
}

\maketitle

\authorrunning{Aa. Sandqvist et al.}

\section{Introduction}

The central region of the Galaxy has been extensively studied at
wavelengths between the near infrared and the radio portions of the
spectrum (see reviews by Morris \& Serabyn \cite{mor96}; Mezger et
al. \cite{mez96}) as well as at $\gamma$-ray and X-ray wavelengths. The 
molecular clouds dominate the interstellar medium in the inner 500 pc
($1\degr  = 150$\ pc) of the Galaxy and the density of molecular 
clouds is far higher in this region than in any other part 
of the Galaxy. Although it represents less than 0.2\%\  of the Galactic disk by
volume, nearly 10\%\  of the total Galactic molecular mass is found
here. A dominant feature in this region is the inclined Expanding Molecular
Ring (EMR, e.g. G\"usten \cite{gus89}). Another feature closer to the
Centre is Sgr B2 which is the most prominent and massive concentration
of molecular gas (GMC) and star formation in the entire
Galaxy. Neufeld et al. (\cite{neu00}) have observed both \htmo\ and
\htio\  towards this source using the Submillimetre Wave Astronomy Satellite
(SWAS). A dust ridge connects Sgr B2 to the regions closer to the
Centre (Lis \& Carlstrom \cite{lis94}).  
The very central Sgr A Complex consists of a nonthermal shell component,
Sgr A East, and a thermal component, Sgr A West. The source Sgr A
West, with its ``mini-spiral arms'', consists of infalling gas
(Killeen \& Lo \cite{kil89}) and contains in its innermost regions the
unique nonthermal radio source Sgr A*, which is the manifestation of a
$2.6 \times 10^6$ M$_\odot$ black hole in the centre of the Milky Way
system (Eckart \& Genzel \cite{eck96}, Sch\"odel et al. \cite{sch02}).   

The molecular complex associated with \sgra\ consists predominantly of
a molecular belt comprising the ``+50 \kms\ cloud'' (M--0.02--0.07),
the ``+20 \kms\ cloud'' (M--0.13--0.08), and the Circumnuclear Disk
(CND) which surrounds \sgrawest\ and has a rotational velocity of the
order of 100 \kms\ in the same direction as the rotation of the
Galaxy. These warm and high-density Galactic Centre  molecular 
clouds are intimately entwined and interact with the continuum
complex described above (Sandqvist \cite{san89} - H$_{2}$CO; Zylka et
al. \cite{zyl90}, \cite{zyl96}; Serabyn et al. \cite{ser94}, Ojha et
al. \cite{ojh01} - \CI; Lindqvist et al. \cite{lin95} - C$^{18}$O, HNCO; Pak et
al. \cite{pak96} - H$_{2}$). All these structures, together with many 
more, are parts of a mechanism complex involving shocks, magnetic
fields and strong UV radiation fields, and may thus function as prime
candidates for \hto\  observations with the Odin satellite. 

\section{Observations}

Odin is a submillimetre/millimetre wave spectroscopy astronomy and
aeronomy satellite, launched on 20 February 2001 from Svobodny,
Russia in far-eastern Siberia. It has a 1.1-m high-precision telescope
with a beam efficiency of about 90\%\ and beamwidths of $2'.1$ and $9'.5$ at
submm and mm wavelengths, respectively. Its pointing uncertainty is
$<10$\as . There are four cryo-cooled submm receivers tunable in
the frequency range of 486 -- 580 GHz with a single sideband temperature
of $\approx 3000$\,K. A cryo-cooled HEMT receiver is tuned to 119 GHz
and has a single sideband temperature of $\approx 600$\,K.The backend
spectrometers consist of an acousto-optical spectrometer (AOS) with a total
bandwidth of 1040 MHz and two auto-correlators with bandwiths in the
range of 100 -- 800 MHz, corresponding to velocity resolutions of
0.08 -- 1.0 \kms. The satellite is described in detail by Frisk et
al. (\cite{fri03}) and the receiver calibration by Olberg et
al. (\cite{olb03}).

Three positions towards Sgr A have so far been observed with Odin,
namely \sgrastar\ with the CND, the +20 \kms\ molecular cloud and the
+50 \kms\  molecular cloud. The coordinates of the observed positions
are given  in Table 1. Observations have been made in the spectral
lines of 119-GHz O$_{2}$, 487-GHz O$_{2}$, 492-GHz \CI, 548-GHz \htio,
557-GHz \htmo, and 576-GHz ($J=5-4$) CO. However, only the data for
\htmo\ and \htio\ have been fully calibrated and reduced so far and they are
presented in Sect. 3. The data for the other spectral lines will be
presented in a subsequent paper.

\begin{table}
\caption{Observed Positions in the Galactic Centre \sgra\ region}
\begin{flushleft}
\begin{tabular}{lll}
\hline\noalign{\smallskip}
      &  $\alpha$(1950.0)      &   $\delta$(1950.0) \\
\hline\noalign{\smallskip}
\sgrastar\  Circumnuclear Disk & 17$^{\rm h}$42$^{\rm m}$29$^{\rm
s}$.3  &  $-28\degr$$59'$18\as \\
\sgra\  +20 \kms\  Cloud  & 17$^{\rm h}$42$^{\rm m}$29$^{\rm s}$.3
&  $-29\degr$$02'$18\as \\
\sgra\  +50 \kms\  Cloud  & 17$^{\rm h}$42$^{\rm m}$41$^{\rm s}$.0
&  $-28\degr$$58'$00\as \\
\noalign{\smallskip}\hline\end{tabular}
\end{flushleft}
\end{table}   

Two observing methods have been employed with Odin. One method is 
Dicke-switching against one of two sky horns with beamwidths of
4\degr.4, displaced 42\degr\  from the main beam. \sgrastar\
CND was observed this way in the \htmo \ line during October 2001. In
order to improve the baselines, full-orbit observations were made of
an empty reference OFF-position at $\alpha$(1950.0)=17$^{\rm
h}40^{\rm m}$26$^{\rm s}$.8, $\delta$(1950.0)=
$-28\degr35'$04\as\ every  second 
orbit (observing period of 60 minutes). The total ON-position time for
the \htmo\ line observations was eight orbits (480 minutes). The
other observing method was total-power position-switching to the above
reference position with a duty cycle of 120 seconds. This was used for \htmo\ 
observations of the  +20 \kms\  (18 orbits) and the +50 \kms\ (27
orbits) clouds and for \htio\ observations of the \sgrastar\
CND-region (58 orbits) during April/May 2002. The AOS was used for all
the \hto\  observations, which results in a velocity resolution of 0.54
\kms\ and a total velocity coverage of 560 \kms\ in the line profiles.

\section{Results}

\begin{figure*}
  \resizebox{\hsize}{!}{\rotatebox{0}{\includegraphics{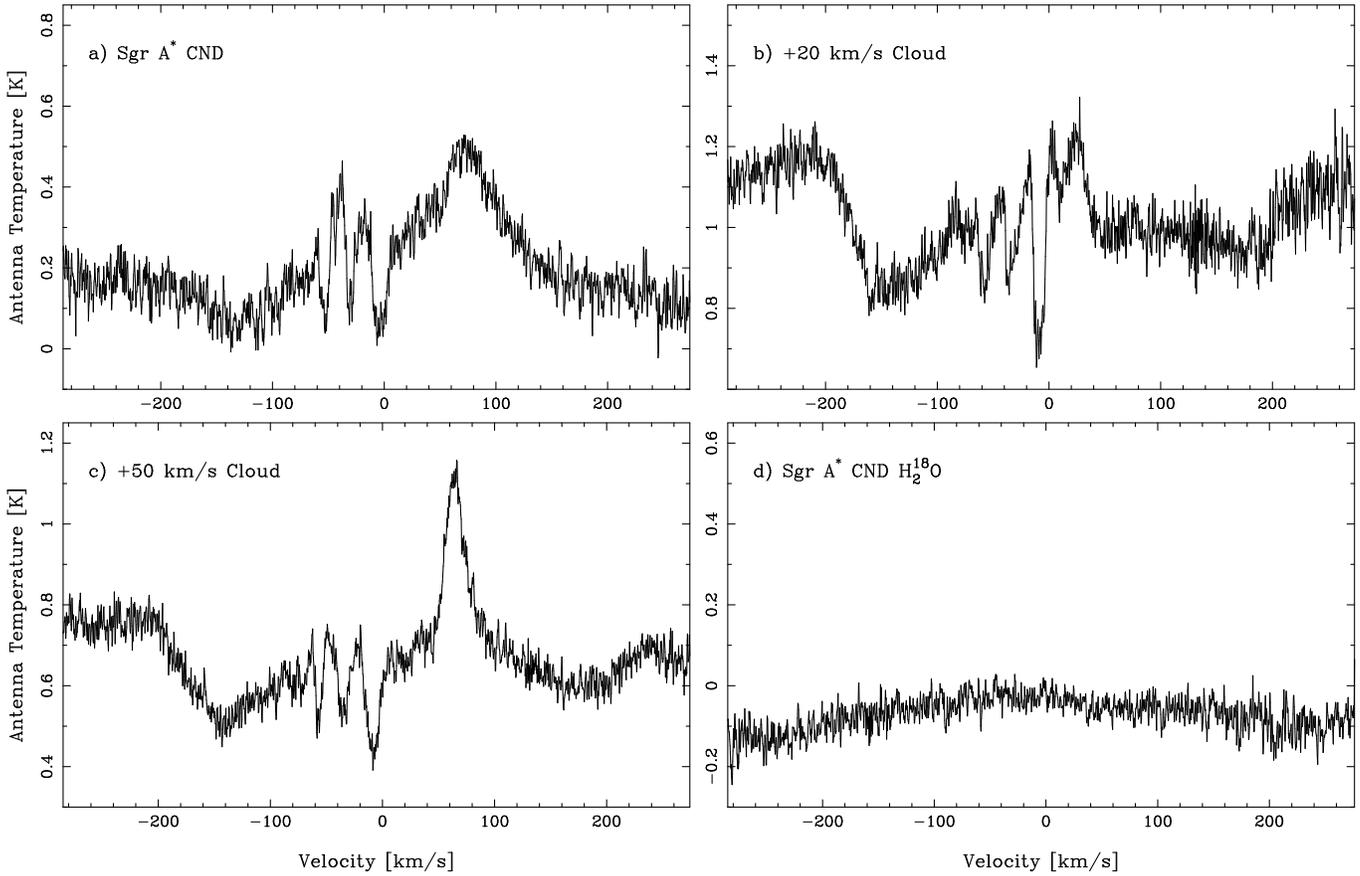}}}
  \caption{The 557-GHz \htmo\  line profiles observed towards {\bf a)}
  the \sgrastar\ Circumnuclear Disk, the {\bf b)} +20 and {\bf c)} +50
  \kms\ clouds, and {\bf d)} the 548-GHz \htio\  isotope line profile
  observed towards the \sgrastar\ Circumnuclear Disk}
  \label{1}
\end{figure*}

\begin{figure*}
  \resizebox{\hsize}{!}{\rotatebox{0}{\includegraphics{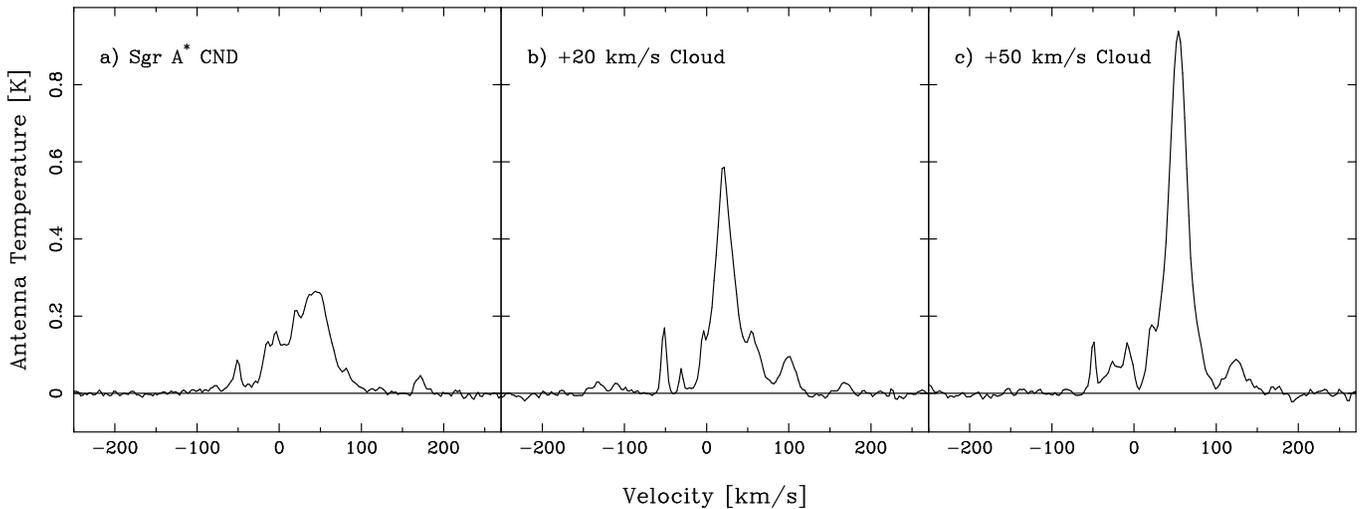}}}
  \caption{The SEST \ciso\ (1--0) survey Odin-beam convolved profiles
  towards {\bf a)} the \sgrastar\ Circumnuclear Disk, the {\bf b)} +20
  and {\bf c)} +50 \kms\ clouds}   
  \label{2}
\end{figure*}

Strong emission and absorption lines have been observed in the \htmo\ 
line at all three \sgra\ positions. However, no spectral line features
can be detected in the \htio\ line down to the rms noise limit of
$\approx 0.02$\, K. The three 557-GHz \htmo\  line profiles observed
towards \sgrastar\ CND, the +20 \kms\ cloud and the +50 \kms\ cloud,
are presented in Figs. 1a-c, and the smooth featureless ($<0.02$\, K rms)
profile of the 548-GHz \htio\ line towards \sgrastar\ CND is shown in
Fig. 1d. This last profile gives an indication of the high quality of the
baselines in our broad-band observations, which is important when
judging the reality of the many emission and absorption features in
the \htmo\ profiles. The intensity scale has not
been corrected for the main beam efficiency ($\approx$ 0.9) in these
four profiles. Furthermore, no baselines have been subtracted. The
intensity thus includes the presence of the background continuum
emission, although the uncertainty of this level is not yet
determined. The relative continuum intensities for the \htmo\
observations conform qualitatively to the continuum level
expected from an interpolation to the \hto\ frequency of the 800 and
350 $\mu$m maps by Lis \& Carlstrom (\cite{lis94}) and Dowell et
al. (\cite{dow99}), respectively. However, it seems that the continuum
level obtained with the Dicke-switching method (Fig. 1a) agrees better
with the interpolated results (see also Sect. 4) than the
position-switching method (Figs. 1b-d). For the sake of comparison
with other spectral lines, we have chosen the data from the SEST
\ciso\ (1 -- 0) survey of the Galactic Centre by Lindqvist et
al. (\cite{lin95}). The \ciso\ profiles resulting for the three \hto\
positions from a convolution of the SEST map spectra to a resolution
of $2'$ (corresponding to the Odin beam size) are shown in Fig.~2.

\section{Discussion}

A Gaussian analysis has been performed on the \sgrastar\ CND
\htmo\  profile using four absorption components and two emission
components (Fig. 3). The continuum emission was first subtracted out
by fitting a linear baseline to the outermost channels on either side
of the profile. The Gaussian analysis results are given in Table 2.

\begin{figure}
  \resizebox{\hsize}{!}{\rotatebox{0}{\includegraphics{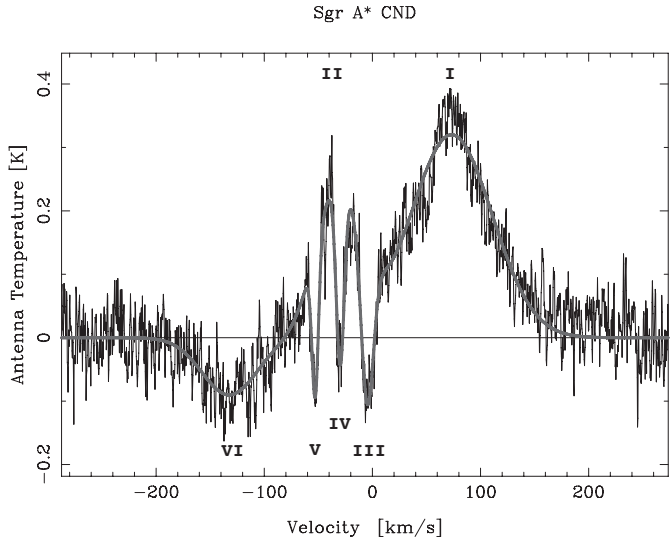}}}   
  \caption{Gaussian analysis performed on the 557-GHz \htmo\ 
  line profile observed towards the \sgrastar\ Circumnuclear Disk}
  \label{3}
\end{figure}

\begin{table}
\caption{Gaussian components of the \sgrastar\  Circumnuclear Disk 557-GHz \htmo\  line profile}
\begin{flushleft}
\begin{tabular}{ccccc}
\hline\noalign{\smallskip}
          & Velocity        &  $T^{*}_{\rm A}$  & Halfwidth & Source        \\
          & (\kms)          & (K)         & (\kms)    &               \\
\hline\noalign{\smallskip}
I         &  $+73.2$        &  $+0.32$    & 88.5      & CND           \\
II        &  $-31.6$        &  $+0.24$    & 47.9      & CND           \\
III       &  $-4.8$         &  $-0.24$    & 13.4      & Local Sgr Arm     \\
IV        &  $-30.2$        &  $-0.25$    & 11.0      & $-30$\ \kms\ Arm\\
V         &  $-53.5$        &  $-0.21$    & 8.1       & 3-kpc Arm     \\
VI        &  $-132.2$       &  $-0.09$    & 60.0      & EMR           \\
\noalign{\smallskip}\hline\end{tabular}
\end{flushleft}
\end{table}   

The first two components, I and II, both seen in emission, are
believed to originate in the rapidly rotating CND. The northeastern part of
the CND is receding and the southwestern part approaching, which gives
the asymmetric, somewhat double-peaked line profile structure. The
2.1-arcmin beam of Odin encloses fully the CND and the resulting
velocity structure of the profile is reminiscent of that seen in many
other molecular lines (see e.g. HCO$^{+}$ ($1-0$) - Linke et
al. \cite{lin81}; HCO$^{+}$ ($3-2$) - Sandqvist et 
al. \cite{san85}; H$_{2}$CO ($2-1$) and CS ($5-4$) - Sandqvist
\cite{san89}; CO ($4-3$) - White \cite{whi96}). 

Three narrow \hto\  absorption components, seen at
velocities near --5, --30 and --53 \kms, are observed at all three
positions and are well-known Galactic spiral arm features, which were
first identified in early 21-cm \HI \ observations. They originate
along the line of sight crossing the so-called Local Sgr, --30 \kms \ and
3-kpc spiral arm structures. 

From the two submillimetre continuum maps discussed in Sect. 3,
we find that the 350$\mu$m:800$\mu$m flux ratios (on a 30\as\ scale)
for all three positions are about 17 - 18 (corresponding to a spectral
index of about 3.5). From the 800 $\mu$m map we estimate the flux
densities on a $2'$ scale to be 160, 300, and 250 Jy, for the
\sgrastar\ CND, +20 and +50 \kms\ cloud positions, respectively. At
557 GHz (538 $\mu$m) and with a conversion factor of 4100 Jy/K (based
on a theoretical $\eta_{\rm a}=0.7$ for Odin) we estimate continuum
levels of 0.16, 0.29, and 0.24 K in our three positions. In the +20
\kms\ cloud profile the deepest absorption is about 0.4 K which is
significantly deeper than our estimated continuum level of 0.29 K.

The three distinct and rather narrow absorption
features (III, IV and V) are present in the spectra at all three
positions (see Figs. 1a-c). The absorption feature (III) at --5 \kms\ 
appears to be the strongest and, judged by the estimated continuum levels, this
feature has an optical depth of at least one. The absorbing gas in these
three features lies in front of the thermal (and non-thermal) continuum 
sources as well as \hto\  gas seen in emission. Since it is not known
how the foreground gas is distributed with respect to the continuum
sources and the \hto\ gas seen in emission and since our estimates of
the continuum levels are uncertain, we shall refrain from calculating
the optical depths of the absorption features. However, when
estimating the (lower) limits of the water column density we shall
assume that the optical depths are $\geq 1$. 

We can estimate the H$_2$ column densities using the \ciso-profiles in
Fig. 2 and calculating the integrated intensities over the regions
corresponding to the three narrow \hto\ absorptions. The H$_2$ column
densities have been calculated by assuming optically thin  emission,
an excitation temperature of 15 K and a C$^{18}$O abundance of
$2\times 10^{-7}$ with respect to H$_2$ (Frerking et
al. \cite{fre82}). The lower \hto\ column density limits, using the
appropriate line width for each absorption feature and an excitation
temperature of 15 K, have been calculated from the assumption that the
optical depths are $\geq$ 1. The corresponding \hto\ abundance limits
are then obtained by using the H$_2$ column density estimates from the
\ciso\ data. The results are summarized in Table 3.

\begin{table*}
\caption{Abundances in the Local Sgr (III), --30 \kms\ (IV) and 3-kpc (V)
Spiral Arm Features} 
\begin{flushleft}
\begin{tabular}{ccccccc}
\hline\noalign{\smallskip}
Position       &  Feature       &  $I$(\ciso) & $N$(\ciso) & $N$(H$_2$)
           &  $N$[\hto]  &  $X$ [\hto]  \\
  &       & (K\ts\kms)        & (cm$^{-2}$)    &
           (cm$^{-2}$) &  (cm$^{-2}$)   &             \\
\hline\noalign{\smallskip}
\sgrastar\ CND  &     III   &      2.10   &   $2.1\times 10^{15}$ &
           $1.0\times 10^{22}$ & $>2\times 10^{13}$  & $> 2\times
           10^{-9}$ \\ 
  &  IV  &        0.18  &   $1.8\times 10^{14}$ &  $9.0\times 10^{20}$ &
           $>9\times 10^{12}$  & $>1\times 10^{-8}$ \\ 
  &   V   &       0.65  &   $6.5\times 10^{14}$ &  $3.2\times 10^{21}$
           & $>8\times 10^{12}$  & $>2\times 10^{-9}$ \\ 
\noalign{\smallskip}
+20 \kms\ Cloud   &    III   &      1.00  &    $1.0\times 10^{15}$ &  $5.0\times 10^{21}$ & $>1\times 10^{13}$  & $> 2\times 10^{-9}$ \\
  &   IV  &      0.34 &    $3.4\times 10^{14}$ &   $1.7\times 10^{21}$
  & $>1\times 10^{13}$  & $>6\times 10^{-9}$ \\ 
  &   V    &     0.96  &  $9.6\times 10^{14}$ &  $4.8\times 10^{21}$ &
  $>2\times 10^{13}$ &  $>4\times 10^{-9}$ \\ 
\noalign{\smallskip}
+50 \kms\ Cloud   &   III   &     1.10  &    $1.1\times 10^{15}$ &  $5.5\times 10^{21}$ & $>2\times 10^{13}$  & $> 4\times 10^{-9}$ \\
  &   IV   &      0.64  &   $6.4\times 10^{14}$ &  $3.2\times 10^{21}$
  & $>1\times 10^{13}$  & $>3\times 10^{-9}$ \\ 
  &    V   &      0.81  &   $8.1\times 10^{14}$ &  $4.0\times 10^{21}$
  & $>6\times 10^{12}$  & $>2\times 10^{-9}$ \\ 
\noalign{\smallskip}\hline\end{tabular}
\end{flushleft}
\end{table*}

A broad \hto \ absoption component (VI), seen at velocities near
--132 \kms, is also observed in all three positions. This feature
has its origin in the near side of the Expanding Molecular Ring
(EMR). The EMR is a massive $\approx$180 pc molecular ring surrounding
the Galactic Centre and it has been observed in many atomic and
molecular species (e.g. Morris \& Serabyn \cite{mor96}). The near
side is seen in broad absorption lines towards the \sgra\ Complex at
velocities of $\approx-130$ \kms, while the far side is seen only in
emission lines, which towards the \sgra\ Complex have velocities near
+170 \kms. 
 
Now let us turn our attention to the \hto \ profiles observed towards the +20
and +50 \kms \ clouds, presented in Figs. 1b and c. In addition to the
four absorption features discussed above, the profiles are marked by
the characteristic emission component from these molecular clouds at
velocities near  +20 \kms \ and +50 \kms, respectively. Furthermore, a
new molecular feature in the Galactic Centre can now been
identified. It is detected as broad \htmo \  absorption in the 
velocity range of $\approx$ +120 to +220 \kms \  (see Figs. 1b and
c). We shall call this feature the High Positive Velocity Gas (HPVG). This
feature is not seen in the \sgrastar \ CND profile (Fig. 1a), which we
interpret as being due to the background continuum emission seen at
this position being somewhat lower than towards the dust continuum
peak emission from the \sgra \ +20 and +50 \kms \ molecular clouds (see
the 800 and 350 $\mu$m continuum maps of Lis \& Carlstrom
\cite{lis94} and Dowell et al. \cite{dow99}, respectively). However, a
careful study of the \hto \ profile in Fig. 1a (and Fig. 3) may show an
extended very weak emission wing in this HPVG velocity range, so there
may be still HPVG present even towards the \sgrastar \ CND region,
although here the background continuum is too weak to cause visible
absorption. Alternatively, the high positive velocity wing of the CND
emission may mask any HPVG absorption. 

Evidence for the existence of the HPVG in the \sgra\ region, seen in
other spectral lines, is scarce. The HPVG should not be confused with
the molecular gas in the far side of the EMR whose velocity falls
inside the same range but whose emission lines are narrower. Also, the
HPVG is seen in absorption which places it in front of the Galactic
Centre continuum sources and thus it cannot be part of the far side of
the EMR. Moneti et al. (\cite{mon01}) have used the Infrared Space
Observatory (ISO) to obtain mid- and far-infrared \hto\ profiles
towards \sgra. These profiles do indeed show some absorption
components at velocities corresponding to that of the HPVG. Some
evidence for the HPVG may also be present in VLA OH absorption
observations towards the \sgracomp\  by Karlsson et
al. (\cite{kar03}). 

Additional evidence for the HPVG is also apparent in the data of a
new high-resolution \HI\ absorption survey of the Galactic Centre
region performed with the VLA by Dwarakanath, Zhao, Goss and Lang
(2003, in preparation). They have kindly convolved their data with the Odin
beam at our positions and find that (1) towards
the +20 \kms\ cloud there is an  \HI\ absorption with an optical depth of
$\sim 0.03$ at +100 \kms, decreasing to $\sim 0$ around +130 \kms\ 
and a second absorption component centred around +150 \kms\ with an
optical depth of $\sim 0.03$ and a width of $\sim 20$ \kms, and
(2) towards the +50 \kms\ cloud in the velocity range of +100 to +200
\kms\  there is an \HI\  absorption with the optical depth decreasing
monotonically from $\sim 0.04$ at +100 \kms\ to $\sim 0$ at +200 \kms.

Although 58 Odin orbits were dedicated to observing the \htio \ line
towards the \sgrastar\ CND position, no spectral line was detected
(see Fig. 1d). Our non-detection of H$_2^{18}$O towards 
\sgrastar\  CND provides an upper limit on the H$_2$O abundance in the narrow
absorption features. Given the rms noise of 23 mK in the H$_2^{18}$O
spectrum and the estimated continuum level of 0.16 K we find that a 10
\kms\  wide absorption feature of optical depth 0.08 should have been
detected at the 3$\sigma$ level. Using this limit and adopting a
$^{16}$O/$^{18}$O ratio of 500 for this local absorbing cloud
(Wilson \& Rood \cite{wil94}) and  an excitation temperature of 15 
K, we obtain an H$_2$O column density of $5\times 10^{14}$
cm$^{-2}$. Hence, for the Local Sgr Arm absorption, the 3$\sigma$ upper
limit of the H$_2$O abundance becomes $5\times 10^{-8}$, using the
H$_2$ column density of $1.0\times 10^{22}$ in Table 3, while the
lower limit was found to be $2\times 10^{-9}$. The average \hto\
abundance estimated for the foreground gas towards Sgr B2 by Neufeld et
al. (\cite{neu00}) is  $6 \times 10^{-7}$, which is about an order of
magnitude higher than our range towards \sgra. On the other hand, our
range is in better agreement with \hto\ abundances found in giant
molecular cloud cores by Snell et al. (\cite{sne00}) and in a local
diffuse molecular cloud by Neufeld et al. (\cite{neu02}).  

In September/October of 2002, Odin again observed the Galactic
Centre region in the \htio \  line, this time pointing at the +20 and
+50 \kms \ cloud positions. These observations have not yet been calibrated and
reduced. They will be reported in a future paper. 

\begin{acknowledgements}

We should like to thank K. Dwarakanath, J.-H. Zhao, M. Goss and
C. Lang for permission to use some of their VLA \sgracomp\ \HI\ absorption
line results before publication and K. Dwarakanath for making the
Odin-compatible analysis of their \HI\ data. 

\end{acknowledgements}

\end{document}